\documentclass{PoS}

\pdfoutput=1
\usepackage{graphicx}

\title{Opportunities From Precision Flavour Physics}

\ShortTitle{Opportunities From Precision Flavour Physics}

\author{\speaker{Marco Ciuchini}\\
        INFN Sezione di Roma Tre, Via della Vasca Navale 84, I-00146, Roma (Italy)\\
        E-mail: \email{ciuchini@roma3.infn.it}}

\abstract{The possible role of precision flavour physics, and particularly of $B$ physics, in the
next decade is briefly discussed. Few 2--3$\sigma$ deviations from the Standard Model
found in present $B$ data are reviewed as potential forerunners of new physics signals to
be looked for in next-generation experiments. The perspectives for theoretical calculations,
in particular those based on lattice QCD, to match the expected progress in experimental precision are
also presented.}

\FullConference{12th International Conference on B-Physics at Hadron Machines - BEAUTY 2009\\
		 September 07 - 11 2009\\
		 Heidelberg, Germany}

\begin{document}

\section{Introduction}
The role and purpose of flavour physics, and particularly $B$ physics, need to be reassessed after the
end of the $B$ factories and the advent of the LHC era. In the past ten years, $B$ physics focused
on the determination of the Cabibbo-Kobayashi-Maskawa (CKM) matrix~\cite{ckm}
parameters within the Standard Model (SM), trying (successfully) to confirm experimentally the
Kobayashi-Maskawa mechanism
for $CP$ violation. Nowadays this task is accomplished: the phase in the CKM matrix is the dominant source
of $CP$ violation in meson decays. In spite of this success, the SM still gives a phenomenological
description of flavour and $CP$ violation without providing any explanation of these phenomena. A more
fundamental theory of flavour and $CP$ violation beyond the SM and is missing at present.
% ~\footnote{
% Interestingly enough, even without invoking a deeper understanding of flavour and $CP$ violation,
% it is possible to generate the CKM quark mixing entirely through loop effects beyond the SM.
% For a recent discussion, see refs.~\cite{arXiv:0810.1613,arXiv:0905.3130}.}.

In any case, the search for physics beyond the SM through virtual effects in flavour-changing neutral
current (FCNC) and CP-violating processes always remained the goal-of-choice for flavour physics,
taking advantage of the multiple suppression factors (GIM mechanism~\cite{Glashow:1970gm},
weakly-coupled loops, small mixing angles, etc.) that characterize these processes in the SM.
Indeed, already present data put strong bounds on flavour and $CP$ violation beyond the SM.
In particular, kaon data (and to a lesser extent $D$ data~\cite{charm}) require either a
New Physics (NP) scale well beyond the TeV region or a mechanism to suppress the flavour- and
$CP$-violating couplings~\cite{Bona:2007vi}. In other words, sensitivity to the exchange of
particles with masses in the multi-TeV range (and even beyond) and/or possibility to put
non-trivial contraints on the NP Lagrangian characterize flavour physics in general.

As far as $B$ physics is concerned, present data are less constraining, given the larger
experimental errors and the larger SM contributions with their theoretical uncertainties.
For example, the $B_d$ mixing amplitude can still accomodate loop contributions from weak-coupled
particles lighter than 1 TeV~\cite{Bona:2007vi}. Yet $B$ physics has the clear advantage that many
channels sensitive to NP contributions of different origin are available, allowing for
exploring a broad range of NP options. Furthermore, one can argue that NP should
more likely affect the third generation. For instance, in many SUSY-GUT scenarios, large neutrino
mixing angles imply large flavour-changing transitions between the third and the second
generation~\cite{susygut}.

In the next decade, $B$ physics will fully enter the precision measurement era moving from the
``10\%'' to the ``1\%'' typical accuracy and improving experimental sensitivities by an order of
magnitude. The accessible NP scale will be pushed in the multi-TeV region with
the possibility to have NP already discovered by direct searches at the LHC.
In any case, $B$ physics could provide a crucial contribution to the NP search and
characterization. 

Assuming that one order-of-magnitude improvement in experimental precision will allow us to
detect NP contributions to $B$ transtions, a first reasonable question is whether we should
not be start seeing some deviations today. Indeed, few 2--3$\sigma$ deviations from
the SM are found in present $B$ data. They could be the forerunners of actual new
physics signals. In Section~\ref{sec:2} we present the most promising ones.

A second crucial question is whether the theoretical calculation of the Standard
Model amplitudes is precise enough not to hinder NP contributions irrespective of the
experimental achievements. In particular, hadronic uncertainties need to be controlled
at ${\cal O}(1\%)$ level. Although there are ``theoretically clean'' measurements which
can provide unambiguous signals of NP without relying on theoretical inputs,
a systematic study of the NP flavour structure needs a precise determination
of several hadronic parameters. The precision reachable in the next years by non-perturbative methods,
particularly lattice QCD, are critically reviewed in Section~\ref{sec:3} and compared with
the requirements coming from the next-generation experiments in $B$ physics.

\section{Deviations From The Standard Model In \boldmath$B$ Physics}
\label{sec:2}

Overall, flavour data are in excellent agreement with the SM expectation
at the present level of accuracy. Yet there are few processes showing
2-3$\sigma$ deviations which could be the harbringers of NP signals.
Certainly, the importance of these effects should not be overemphasized,
as low-significance deviations could result from statistical fluctuations,
which are expected in the presence of several different measurements.
However, these are processes potentially sensitive to NP contributions.
In particular, the observed deviations are expected in
several interesting NP scenarios. Therefore, while it is premature to
get excited about NP, it is certainly worth keeping an eye on these processes.

The first herald of NP we consider is the phase $2\phi_s$ of the $B_s$ mixing
amplitude. In the SM (with the CKM phase convention), this phase is very small.
It is given by $2\phi_s=-2\beta_s=-0.041\pm 0.004$, where the angle
$\beta_s=\mathrm{arg}\left(-(V_{tb}^*V_{ts})/(V_{cb}^*V_{cs})\right)$
is precisely determined by the Unitarity Triangle (UT) fit~\cite{arXiv:0908.3470}.
A NP amplitude carrying a new CP-violating phase contributing to the $B_s$ mixing
can drastically change this prediction.

\begin{figure}[tb]
\centering
\begin{minipage}[t]{0.52\linewidth}
    \centering
    \includegraphics[width=\textwidth]{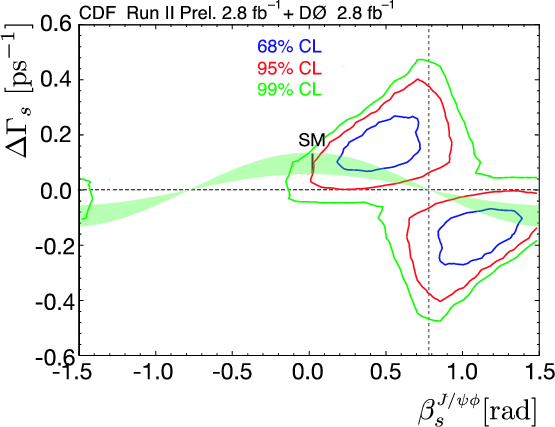} 
    \caption{$68\%$, $95\%$ and $99\%$ confidence regions in the $\beta_s^{J/\psi\phi}$--$\Delta\Gamma_s$
 plane ($\beta_s^{J/\psi\phi}=-\phi_s$) from the preliminary combination of the CDF and D\O\
 results~\cite{jpsiphicomb}.}
\label{fig:jpsiphi}
\end{minipage}
\hspace{0.3cm}
\begin{minipage}[t]{0.44\linewidth}
    \centering
     \includegraphics[width=\textwidth]{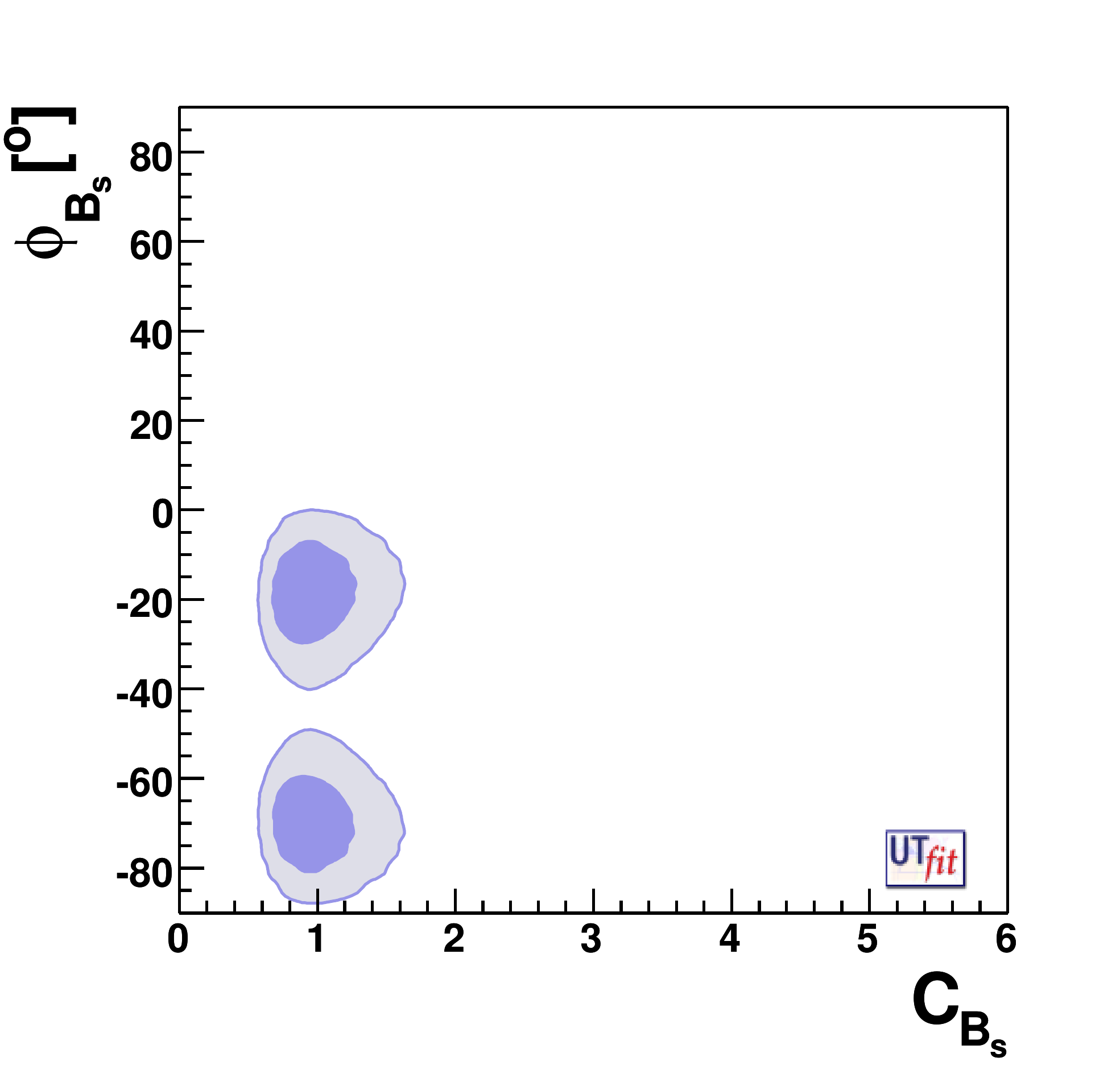} 
    \caption{68\% (dark) and 95\% (light) probability regions in the NP parameter space ($C_{B_s},\phi_{B_s}$).}
\label{fig:cbsphibs}
\end{minipage}
\end{figure}

Recently, the time-dependent angular analysis of $B_s\to J/\psi\phi$ was
measured at the Tevatron~\cite{phisexp,jpsiphicomb}. From this measurement, one can extract a combined
determination of $\Delta\Gamma_s$ and $\beta_s^{J/\psi\phi}$. The angle $\beta_s^{J/\psi\phi}$
is $-\phi_s$ up to doubly Cabibbo-suppressed terms. In the same approximation the
$B_s$ mixing phase vanishes, so that a large $\beta_s^{J/\psi\phi}$ is a clean NP
signal. The combination of the CDF and D\O\ measurement is shown in fig.~\ref{fig:jpsiphi}.
The combined mesurement is found to be compatible with the SM at 2.12$\sigma$.

% \begin{figure}[tb]
%     \centering
%      \includegraphics[width=0.6\textwidth]{phiBs} 
%     \caption{68\% (dark) and 95\% (light) probability regions in the ($C_{B_s},\phi_{B_s}$)-plane.}
% \label{fig:cbsphibs}
% \end{figure}

The UTfit collaboration combined in a Bayesian fit the available results on $B_s\to J/\psi\phi$ from the Tevatron
with other constraints coming from $\Delta\Gamma_s$, the $B_s$ lifetime in flavour-specific final states,
the dimuon charge asymmetry and the semileptonic asymmetry using theoretical correlations among
the observables under the assumption that NP contributions to tree-level dominated processes are negligible.

Using also with the measurement of mass difference $\Delta m_s$, one can find a bound on the NP
contribution to the $B_s$ mixing amplitude. This contribution can be described by the parameters $C_{B_s}$
and $\phi_{B_s}$ defined as
\begin{equation}
C_{B_s}\,e^{2 i \phi_{B_s}} = \frac{\langle B_{d,s} | H_{eff}^{full}| \bar B_{d,s} \rangle}
{\langle B_{d,s} | H_{eff}^{SM}| \bar B_{d,s} \rangle}=\frac{\Delta m_s}{\Delta m_s^\mathrm{SM}} e^{2i(\phi_s+\beta_s)}\,.
\end{equation}
Perfoming a full fit~\cite{arXiv:0905.3747}, one gets the allowed regions in Figure~\ref{fig:cbsphibs}.
The deviation from the SM goes up to 2.9$\sigma$~\cite{arXiv:0905.3747,arXiv:0803.0659}.
Notice that a confirmed evidence
for a new CP-violating phase would rule out the class of models with minimal flavour violation~\cite{mfv},
with intriguing implications for the NP flavour structure, in particular if new particles are found at the LHC.

For these reasons, the $B_s$ mixing phase is bound to play a major role in the indirect searches for NP in
upcoming years. Indeed the NP contribution to this phase could still be much larger than the SM one. If this is the
case, its determination is a null test of the SM, almost free of theoretical uncertainty. A 5$\sigma$ measurement
of $\phi_s$ could even be possible at the Tevatron, if the present central value is confirmed. In any case, this is
one of the flagship measurement of LHCb, which should be sensitive to a value of $\phi_s$ as small as to few degrees.
For such values, however, theoretical uncertainties coming from subleading amplitudes cannot be neglected
anymore and could limit the accuracy achievable on $\phi_s$.

Another NP-sensitive decay showing a potentially interesting deviation from the SM prediction is $B\to\tau\nu$. The
measured branching ratio is~\cite{btaunuexp}
\begin{equation}
BR_\mathrm{exp}(B\to\tau\nu) = (1.73 \pm 0.34)\times 10^{-4}\,.
\label{eq:btnexp}
\end{equation} 
while the SM prediction is given by
\begin{equation}
BR(B \to \tau \nu) = 
\frac{G_{F}^{2}m_{B}m_{\tau}^{2}}{8\pi}\left(1-\frac{m_{\tau}^{2}}
{m_{B}^{2}}\right)^{2}f_{B}^{2}|V_{ub}|^{2}\tau_{B}\,.
\label{eq:btaunu}
\end{equation}
The Fermi constant $G_F$, the $B$ ($\tau$) mass $m_B$ ($m_\tau$) and
the $B$ lifetime $\tau_B$ are precisely measured. The main sources
of theoretical uncertainties are the $B$ meson decay constant
$f_B$ and the CKM matrix element $\vert V_{ub}\vert$. In the following,
$\overline{BR}$ denotes the prediction obtained with eq.~\ref{eq:btaunu}.
It coincides with the SM prediction if $\vert V_{ub}\vert$ and $f_B$
are determined with NP-insensitive methods. Otherwise, the deviation
of $\overline{BR}$ from the measurement shows the presence of a
non-standard contribution to the $B\to\tau\nu$ amplitude within the NP
scenario used to determine $\vert V_{ub}\vert$ and $f_B$.

\begin{table}
 \begin{center}
\begin{tabular}{l|r|r|r|r}
 scenario & $\vert V_{ub}\vert\times 10^4$ & $f_B$ (MeV) &
 $\overline{BR}\times 10^4$ & pull\\\hline 
 no-fit & $36.7\pm 2.1$ & $200\pm 20$ & $0.98\pm 0.24$ & $1.8\sigma$\\
 UUT    & $35.0\pm 1.2$ & $200\pm 20$ & $0.87\pm 0.20$ & $2.2\sigma$\\
 UT     & $35.2\pm 1.1$ & $196\pm 11$ & $0.84\pm 0.11$ & $2.5\sigma$
 \end{tabular}
 \end{center}
\caption{Results for $\vert V_{ub}\vert$, $f_B$, $\overline{BR}$ and
  the pull between $\overline{BR}$ and $BR(B\to\tau\nu)_\mathrm{exp}$ in different
scenarios (see text).}\label{tab:btaunu}
\end{table}

The status of the comparison between the measurement in Eq.~(\ref{eq:btnexp})
and the theoretical prediction $\overline{BR}$ is summarized
in the following Table~\ref{tab:btaunu} from Ref.~\cite{arXiv:0908.3470}.

In the ``no-fit'' scenario, one uses the determination of $f_B$ from
lattice QCD and $\vert V_{ub}\vert$ from the average of inclusive and
exclusive $b\to u$ semileptonic decays (involving theoretical input
from operator product expansion techniques in the inclusive case and
a combination of lattice QCD and QCD sum rules in the exclusive ones).
This scenario uses a NP-independent determination (neglecting presumably
small NP contributions to tree-level semileptonic decays) of the theoretical
inputs and is therefore appropriate to measure generic non-standard contributions
to the $B\to\tau\nu$ decay amplitude. As shown in Table~\ref{tab:btaunu},
the significance of the deviation from the SM, given by the presence of a
non-standard $B\to\tau\nu$ amplitude, is rather small (1.8$\sigma$),
partly due to the uncertainty of the SM prediction which is not negligible
with respect to the experimental error.

However, if we are generically interested in a deviation from the SM, whatever
the origin is, then we can do better exploiting the SM correlations among
flavour observables. Using the SM UT fit, the determination
of both $\vert V_{ub}\vert$ and $f_B$ can be improved~\cite{hep-ph/0606167},
resulting in a more precise SM prediction for $BR(B\to\tau\nu)$ and therefore an
increased sensitivity to possible deviations. Indeed, the UT scenario in Table~\ref{tab:btaunu}
shows that the errors of $\vert V_{ub}\vert$ and $f_B$ are halved by using
the SM UT. As a consequence, the significance of the discrepancy between the measurement and
the prediction of $\overline{BR}$ increases up to 2.5$\sigma$.

Of course, there is a price to pay for the larger significance. We can no longer disentangle
a genuine NP effect in the decay amplitude from a deviation from the scenario adopted in the UT fit.
This a good example of the dual role of flavour physics in NP studies. On the one hand, flavour
physics can be used to look for significant deviations from the SM regardless of the details of
the NP involved. Potentially, it is very effective in accomplish this task as the only requirement
is to compute the flavour observables as precisely as possible in the SM, fully exploiting the SM
correlations to reduce the uncertainties. Indeed, should NP escape detection from direct 
searches at the LHC, flavour physics could still provide evidence that the NP scale is not too far
away. On the other hand, if some information about NP is known, hopefully from high $p_T$ physics, then
the study of flavour physics can provide unique insights on the structure of the underlying NP model.
Admittedly, this is a difficult and long-term program for which we need to rely on theory in
order to control the uncertainties that could reduce the possibility for flavour physics to disentangle
different NP scenarios.

\begin{figure}[tb]
    \centering
\begin{minipage}[t]{0.45\linewidth}
    \centering
    \includegraphics[width=\textwidth]{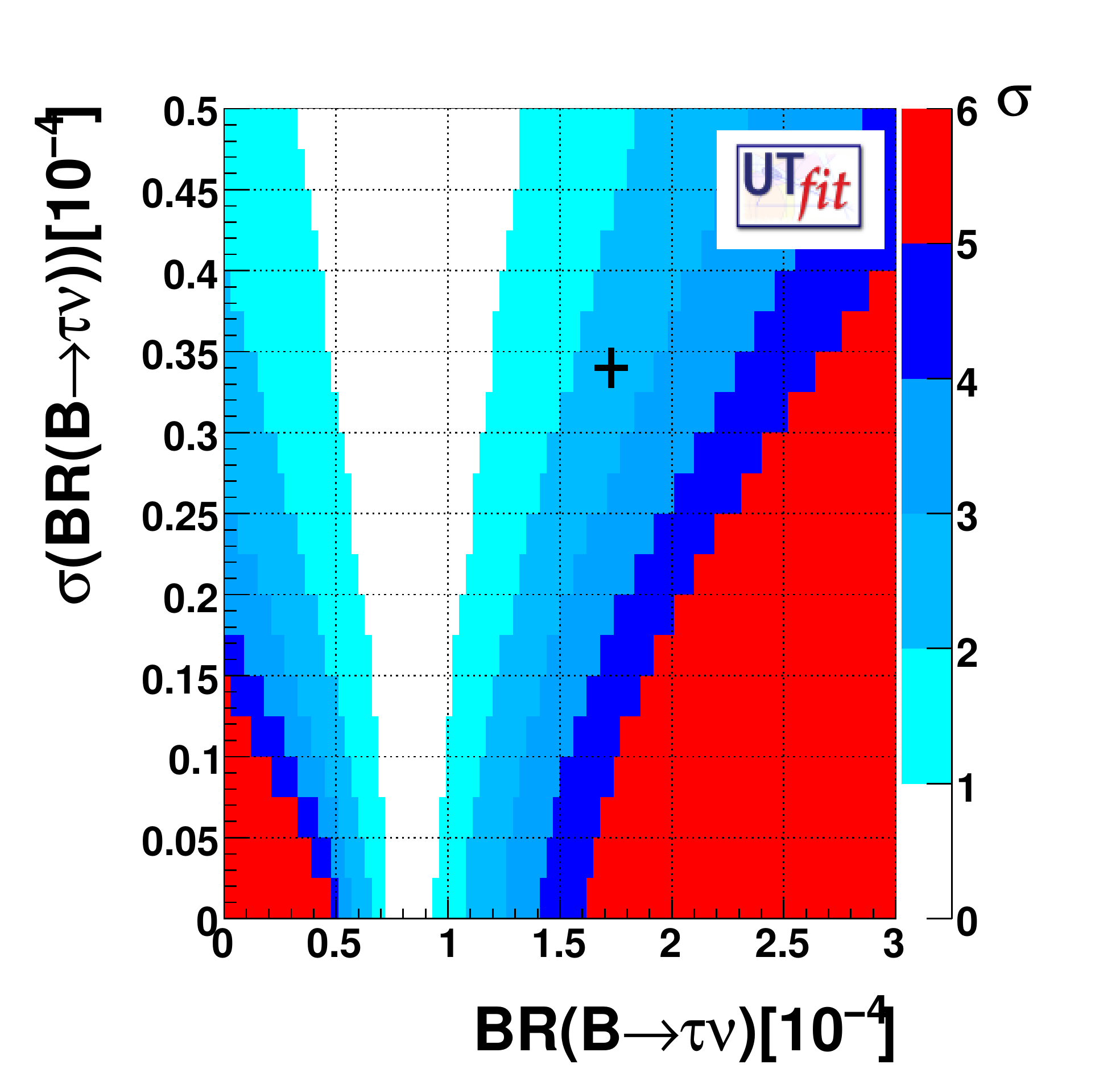} 
    \caption{Compatibility plot for $BR(B \to \tau
      \nu)$. The cross marks the current world average. Colours give
      the agreement (in number of $\sigma$) with the data-driven SM
      prediction.}
\end{minipage}
\hspace{0.3cm}
\begin{minipage}[t]{0.51\linewidth}
     \centering
     \includegraphics[width=\textwidth]{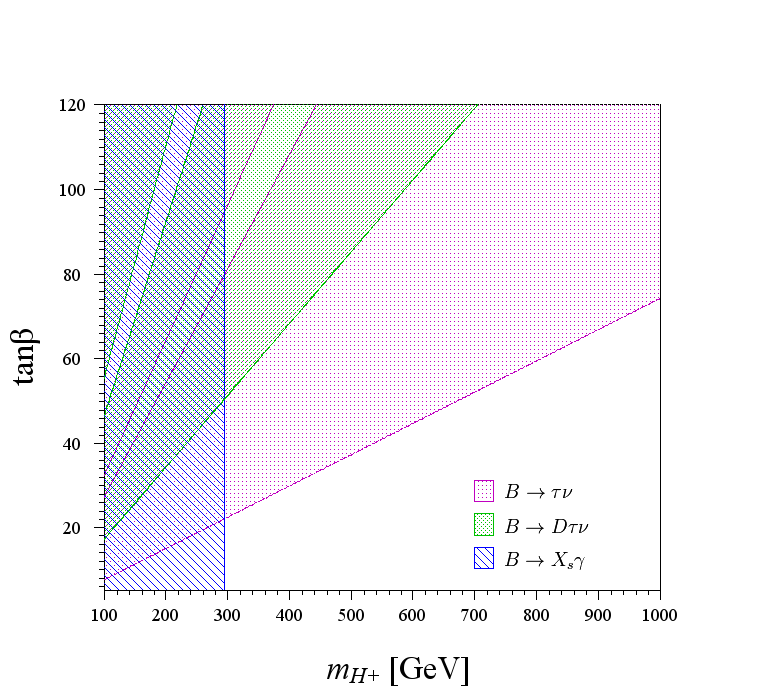} 
     \caption{Regions in the $(m_{H^+},\tan\beta)$ parameter space of
       the 2HDM-II excluded at $95\%$ probability by
       $BR(B\to\tau\nu)$, $BR(B\to D \tau\nu)/BR(B\to D \ell\nu)$ and
       $BR(B\to X_s\gamma)$.\label{fig:tanbmh2d}}
\end{minipage}
\end{figure}

The case marked as ``UUT'' in Table~\ref{tab:btaunu} corresponds to an intermediate
scenario: NP is allowed to enter the UT fit provided it is of minimal flavour violation
type. In this case, correlations are less stringent so that the UT fit decreases only the error of
$\vert V_{ub}\vert$. Thus the prediction of $\overline{BR}$ is not as precise as in the previous case
and the significance of its deviation from the measured value is reduced to 2.2$\sigma$.

% \begin{figure}[tb]
%      \centering
%      \includegraphics[width=0.7\textwidth]{tanbvsmhp_2HDM.png} 
%      \caption{Regions in the $(m_{H^+},\tan\beta)$ parameter space of
%        the 2HDM-II excluded at $95\%$ probability by
%        $BR(B\to\tau\nu)$, $BR(B\to D \tau\nu)/BR(B\to D \ell\nu)$ and
%        $BR(B\to X_s\gamma)$.\label{fig:tanbmh2d}}
% \end{figure}

This is the appropriate case to discuss, as a concrete example, the type-II 2 Higgs Doublet Model
(2HDM-II)~\cite{2hdm}. In this model, one simply obtains
\begin{equation}
BR(B\to\tau\nu) = BR(B\to\tau\nu)_\mathrm{SM}\left(1-\tan^2\beta \frac{m_B^2}{m_{H^+}^2}\right)^2\,,
\label{eq:BR2hdm}
\end{equation}
where $m_{H^+}$ is the mass of the charged Higgs boson and $\tan\beta$ is the ratio of the vacuum expectation
values of the two Higgs doublets. From Eq.~\ref{eq:BR2hdm}, one can see that the charged Higgs contribution
typically suppresses the SM prediction of $BR(B\to\tau\nu)$. As the experimental average is larger than the
SM prediction, a bound on the ratio $\tan\beta/m_{H^+}$ is obtained. Figure~\ref{fig:tanbmh2d} shows the bounds
in the $(m_{H^+},\tan\beta)$ parameter space induced by the constraints coming from the measurements of
$BR(B\to\tau\nu)$, $BR(B\to D \tau\nu)/BR(B\to D \ell\nu)$ and $BR(B\to X_s\gamma)$ are shown.
Combining these three measurements, one gets the following bound with $95$\% probability~\cite{arXiv:0908.3470}:
\begin{equation}
  \label{eq:2HDMbound}
  \tan\beta<7.4 \frac{m_{H^+}}{100\, \mathrm{GeV}}\,,
\end{equation}
together with $m_{H^+}>295$ GeV. Following Ref.~\cite{arXiv:0908.3470}, one can make a prediction for
$BR(B_s\to\mu^+\mu^-)$, another flagship measurement of LHCb, finding
\begin{eqnarray}
  \label{eq:BSllbound}
  BR(B_s\to\mu^+\mu^-) &=& (4.3 \pm 0.9) \times 10^{−9} \\
  &&([2.5, 6.2] \times 10^{−9}~ @95\%~\mathrm{probability}) \nonumber
\end{eqnarray}
The $95\%$ upper bound in Eq.~(\ref{eq:BSllbound}) is stronger than the present upper limit from
direct searches at the Tevatron, $BR(B_s\to\mu^+\mu^-)<5.8\times 10^{-8}$ at $95\%$ C.L.~\cite{bsmm}.
Yet, the latter will be certainly improved by LHCb
which is expected to probe this branching ratio down to the SM value,
$BR(B_s\to\mu^+\mu^-)_\mathrm{SM}=(3.7 \pm 0.5) \times 10^{-9}$. 

\begin{figure}[tb]
 \centering
\begin{minipage}[t]{0.48\linewidth}
     \centering
     \includegraphics[width=\textwidth]{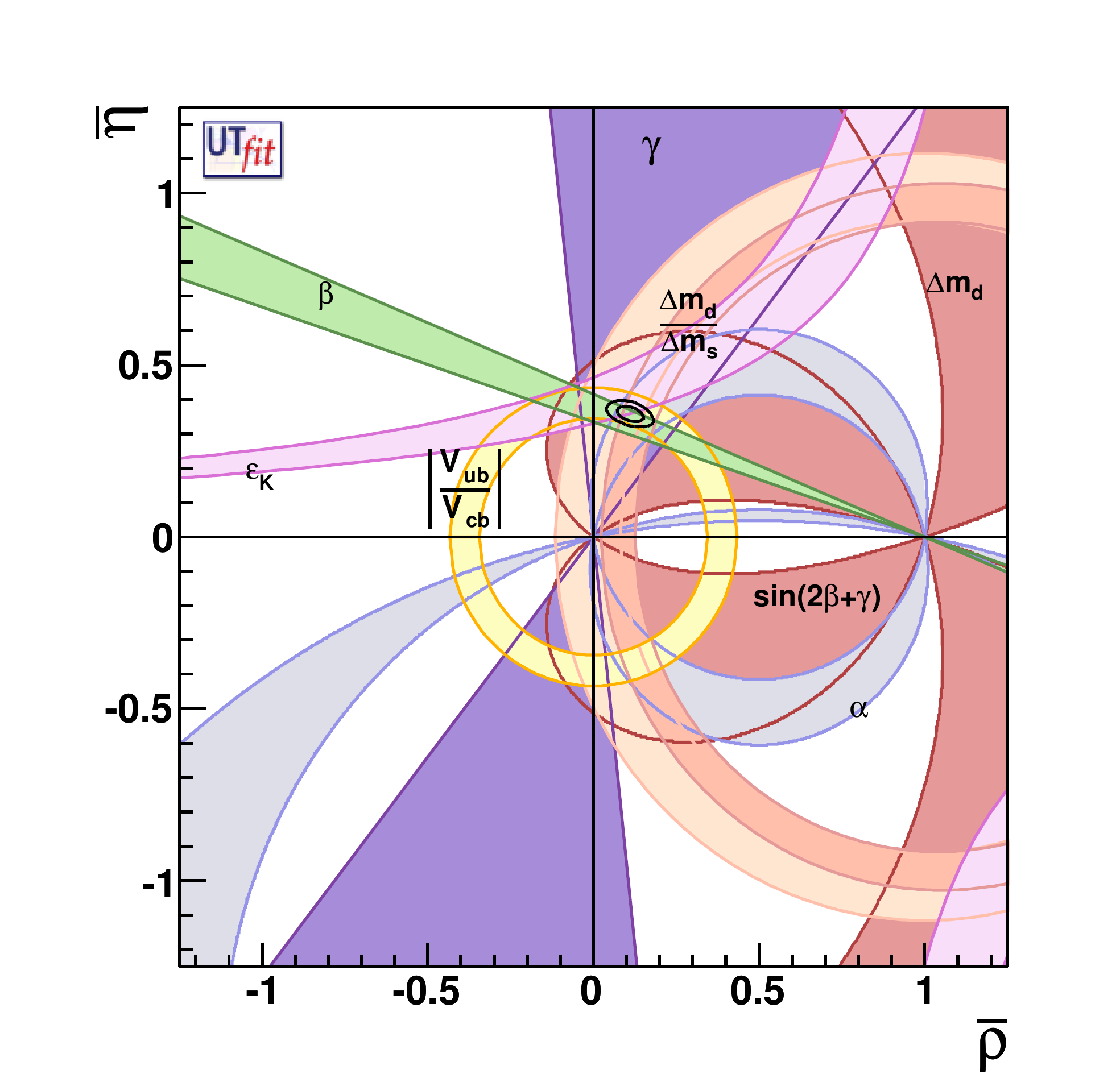} 
    \caption{Result of the UT fit within the SM. The contours display the 68\% and 95\% probability
     regions selected by the fit in the $(\bar \rho, \bar \eta)$ plane. The 95\% probability regions
     selected by the single constraints are also shown.\label{fig:smut}}
\end{minipage}
\hspace{0.3cm}
\begin{minipage}[t]{0.48\linewidth}
     \centering
     \includegraphics[width=\textwidth]{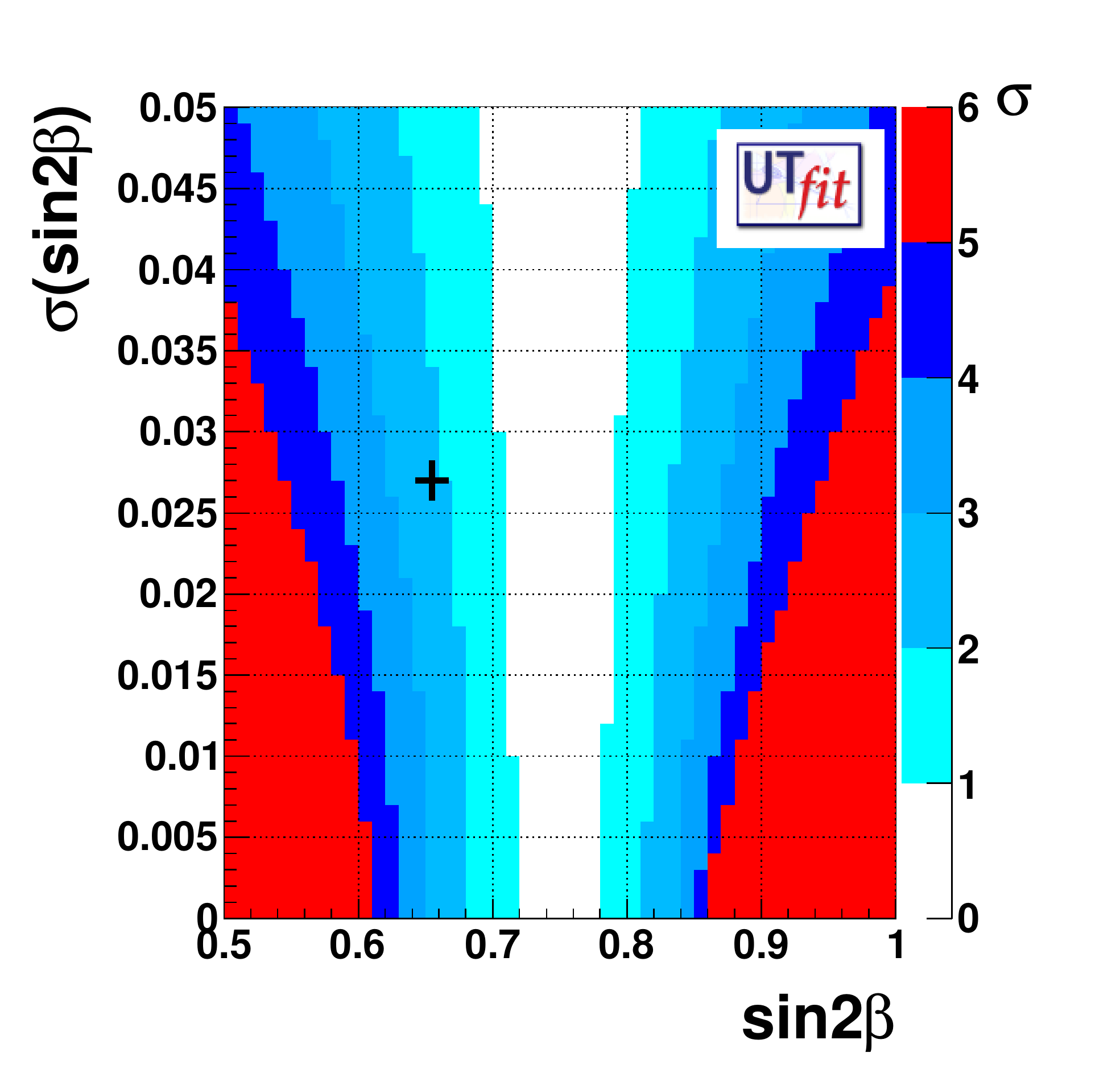} 
    \caption{Compatibility plot for $\sin2\beta$. The cross marks the current world average. Colours give
      the agreement (in number of $\sigma$) with the data-driven SM
      prediction.\label{fig:sin2b}}
\end{minipage}
\end{figure}

The last deviation from the SM we want to discuss is located in the UT fit. The present result of the SM UT fit is
shown in Figure~\ref{fig:smut} together with the 95\% probability regions selected by the various
constraints~\cite{arXiv:0908.3470}. Clearly some constraints, in particular those coming from the measurements of
the CP-violating parameters $\sin2\beta$ and $\varepsilon_K$, do not perfectly overlap. This can be seen as a
deviation of the measured $\sin2\beta$ from the value selected by the other constraints (alternatively, one
can consider the difference between the fit and the lattice determinations of the bag parameter $B_K$ entering
$\varepsilon_K$). The significance of this deviation is shown in Figure~\ref{fig:sin2b}. The
value $\sin2\beta_\mathrm{J/\psi K}=0.655\pm 0.027$, measured from time-dependent $CP$ asymmetry in
in the decays $B\to J/\psi K$, deviates from the UT fit determination (not including
$\sin2\beta_\mathrm{J/\psi K}$) $\sin2\beta_\mathrm{fit}=0.751\pm 0.035$ at about 2.2$\sigma$.

% \begin{figure}[tb]
%      \centering
%      \includegraphics[width=0.6\textwidth]{pull_sin2bfit3} 
%     \caption{Compatibility plot for $\sin2\beta$. The cross marks the current world average. Colours give
%       the agreement (in number of $\sigma$) with the data-driven SM
%       prediction.\label{fig:sin2b}}
% \end{figure}

Actually, $\sin2\beta_\mathrm{J/\psi K}$ has always been somewhat smaller than $\sin2\beta_\mathrm{fit}$
and this ``tension'' was already observed some years ago. However, this effect reduced with time and furthermore
it could be almost entirely ascribed to theoretical uncertainties in the determination of $\vert V_{ub}\vert$ from
inclusive decays~\cite{hep-ph/0606167}.
Recently this issue was revived by the combined effect of the increasing precision of $B_K$ computed on the
lattice and a reanalysis of the theoretical expression of $\varepsilon_K$~\cite{arXiv:0803.4340,arXiv:0805.3887}.
In Ref.~\cite{arXiv:0805.3887} it was pointed out that some approximations commonly used is the formula of
$\varepsilon_K$ are no longer justified. In particular, the theoretical expression of $\varepsilon_K$ used a fixed
value of $\pi/4$ for the phase $\delta_0$ of the $K\to\pi\pi$ $\Delta I=1/2$ amplitude $A_0$ and neglected a
subleading (in the CKM convention) term proportional to Im$A_0$/Re$A_0$.  Indeed, using the measured value of
$\delta_0$ and the value of Im$A_0$/Re$A_0$ estimated using $(\varepsilon^\prime/\varepsilon)_K$, the theoretical
prediction of $\varepsilon_K$ is reduced by about 8\% and the corresponding constraint in the $(\bar\rho, \bar\eta)$
plane is shifted upward.

This is a good example of how indirect NP searches should work. With the reduction of experimental and theoretical
uncertainties, the sensitivity to NP contributions increases and discrepancies could appear in observables
previously compatible with the SM. Clearly, in the case of $\varepsilon_K$, it is premature to draw
any conclusion: not only the statistical significance is still low, but also there are other neglected
terms in the theoretical expression of $\varepsilon_K$, such as power-suppressed terms of ${\cal O}(m_K/m_c)^2$,
which could be non-negligible as well.

\section{The Future Of Hadronic Uncertainties}
\label{sec:3}

\begin{figure}[tb]
     \centering
     \includegraphics[width=0.5\textwidth]{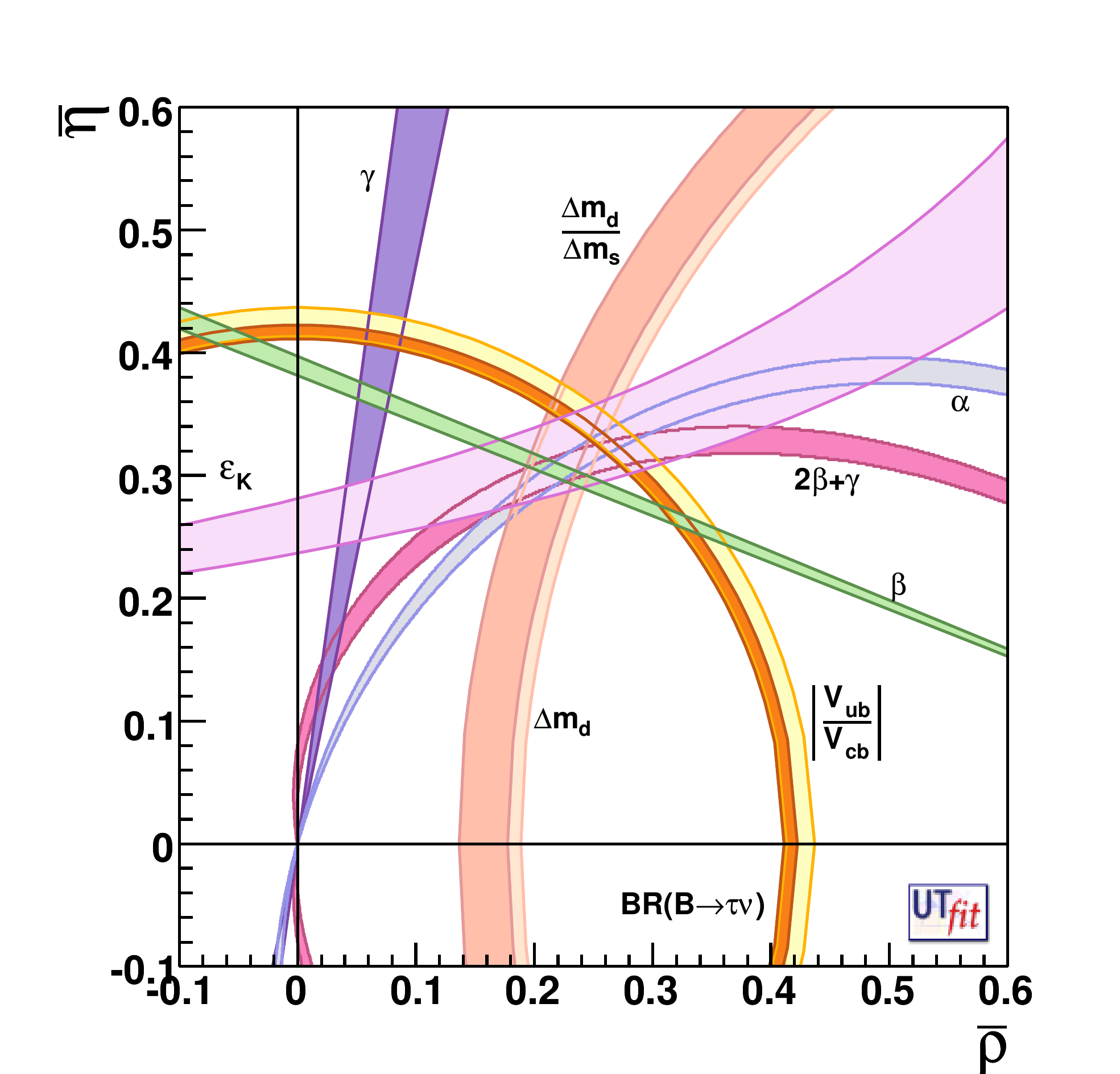}
    \caption{UT fit within the SM extrapolated using expected results from next-generation flavour experiments and
     future lattice QCD calculations (see text). Central values of the constraints are chosen from the present UT
     fit. The contours display the 68\% and 95\% probability
     regions selected by the fit in the $(\bar \rho, \bar \eta)$-plane. The 95\% probability regions
     selected by the single constraints are also shown.\label{fig:ut2015}}
\end{figure}

There are flavour observables, such as $\phi_s$ in the previous section, wich are sensitive to NP irrespective
of theory uncertainties (at least to some extent). In other cases, as for $BR(B\to\tau\nu)$, the theory error
can be reduced with the help of additional measurements. Yet, in order to exploit the full power of flavour
physics for NP search and, even more, characterization, improved theory predictions are essential. In
particular, hadronic uncertainties need to be controlled with an unprecendented accuracy.
We can consider for example the UT analysis. An extrapolation of the UT fit using the expected precision
of next-generation flavour experiments is shown in Figure~\ref{fig:ut2015}~\cite{Ciuchini:2007zz,arXiv:0710.3799,arXiv:0709.0451}.

This Figure is obtained using the extrapolation of lattice results in Table~\ref{tab:lattice}
taken from the Appendix of Ref.~\cite{arXiv:0709.0451}. The study presented there showed that the lattice
predictions for the hadronic parameters entering the UT fit could reach a ${\cal O}(1\%)$ precision 
as the available computer power will reach few PFlops, namely within 5-10 years.

{ \setlength{\tabcolsep}{2pt}  % special column width for wide table
\begin{table}[tbh]
\begin{center}
\begin{tabular}{lccc||c||cc}  
\hline
\hline
Measurement & 
\begin{tabular}{c} Hadronic \\ Parameter \end{tabular} &  
\begin{tabular}{c}  Status \\ End 2006 \end{tabular} & 
\begin{tabular}{c} 6 TFlops \\ (Year 2009) \end{tabular}&
\begin{tabular}{c} {\bf Status } \\ {\bf End 2009} \end{tabular}&
\begin{tabular}{c} 60 TFlops \\ (Year 2011) \end{tabular}&
\begin{tabular}{c} 1-10 PFlops \\ (Year 2015) \end{tabular} \\ \hline
$K \to \pi\, l\, \nu$   & $f_+^{K\pi}(0)$
& 0.9\,\% & 0.7\,\% & 0.5\,\% & 0.4\,\% & $<0.1\,\%$ \\
$\varepsilon_K$ & $\hat B_K$
& 11\,\%  &  5\,\%  &   5\,\%  & 3\,\%  & 1\,\% \\
$B \to l\, \nu$  & $f_B$
& 14\,\% & 3.5-4.5\,\% &  5\,\%  & 2.5-4.0\,\% &  1.0-1.5\,\% \\
$\Delta m_d$ & $f_{Bs}\sqrt{B_{B_s}}$ 
& 13\,\% & 4-5\,\% &  5\,\%  & 3-4\,\% & 1-1.5\,\% \\
$\Delta m_d/\Delta m_s$ & $\xi $
& 5\,\% & 3\,\% &   2\,\%  & 1.5-2\,\% & 0.5-0.8\,\% \\
$B\to D/D^*\,l\,\nu$& ${\cal F}_{B\to D/D^*}$
& 4\,\%  &  2\,\%  &  2\,\%  &  1.2\,\% & 0.5\,\% \\
$B\to \pi/\rho \,l\,\nu$ & $f_+^{B\pi},\ldots$
& 11\,\% & 5.5-6.5\,\% &  11\,\%  & 4-5\,\% & 2-3\,\% \\
$B\to K^*/\rho \,(\gamma, l^+l^-)$ & $T_1^{B\to K^*/\rho}$
& 13\,\% & ------ &   13\,\%  & ------ & 3-4\,\% \\
\hline
\end{tabular}
\end{center}
\caption{Prediction of the accuracy on the lattice QCD
determinations of various hadronic parameters from the Appendix of Ref.~\cite{arXiv:0709.0451}.
The fifth column has been added following the update of Ref.~\cite{lubicz}.}
\label{tab:lattice}
\end{table}

Comparing the fourth and fifth columns of Table~\ref{tab:lattice}, it is reassuring to see that
the results obtained after three years followed the extrapolations of Ref.~\cite{arXiv:0709.0451}
quite well, with only one exception due to the lack of new studies for that parameter.
Lattice QCD seems able to keep up with the upcoming progress expected from next-generation flavour
physics experiments, allowing to fully exploit their potential in searching and constraining NP.

However, lattice QCD is unable compute hadronic parameters relevant for several important observables.
For example, inclusive modes in semileptonic $B$ decays rely on the heavy quark expansion. For this class
of decays, it seems unlikely that theory alone could be able to reach the required accuracy, due
to the difficulties related to computing matrix elements of subleading operators. On the other hand,
data can be used to control the unknown terms. The perspectives for reaching a target accuracy of
few percent are encouraging~\cite{ligeti}.

Finally, two-body non-leptonic $B$ decays looks more problematic as far as hadronic uncertainties are concerned.
When applicable, predictions are based on some formulation of factorization valid in the infinite mass
limit~\cite{factorization}. Also in these cases, accuracy is limited by uncalculable power-suppressed terms.
The strategy of using data typically requires the use of flavour symmetries, which however have a
theoretical uncertainty themselves which cannot be precisely estimated. For these reasons, precision
flavour physics with non-leptonic decays is particularly difficult. Each decay mode requires a careful
assessment of the involved hadronic uncertainties which otherwise could mask or fake a NP signal.

\section{Conclusions}
\label{sec:end}
In these proceedings, we have discussed few processes which already show a 2-3$\sigma$ deviation
from the SM which could be the forerunners of new physics signal for the next-generation flavour experiments.
Indeed, the era of precision flavour physics is already starting with LHCb at the LHC and will likely continue
in the next decade with a super flavour factory~\cite{arXiv:0710.3799}. Identification and characterization of
new physics in the TeV and multi-TeV regions are the achievable goals, exploiting the reach phenomenology of
$B$, $D$ and $K$ decays, together with lepton flavour violation, to probe many different NP scenarios. To fully
pursue this program, a substantial reduction of theoretical uncertainties would be of great help. 
Theory will have hard (but stimulating) time trying to keep up with the expected experimental progress
on a 10-year scale. Yet, as the case of lattice QCD seems to prove, the mission is not impossible.
Flavour physics is ready to move to the precision era and play its part in challenging the SM harder and
harder. 

\section*{Acknowledgements}
We warlmy thank the Participants and the Organizers for the interesting talks, the great atmosphere,
the enjoyable concert and the fantastic walks on the Philosophenweg.

\end{document}